\documentstyle[epsf,amsfonts,amssymb,eqsecnum,aps,twocolumn]{revtex}
\begin{document}
\draft
\wideabs{
\title{Fluids of platelike particles near a hard wall}
\author{L. Harnau and S. Dietrich}
\address{Max-Planck-Institut f\"ur Metallforschung, 
              Heisenbergstrasse 1, D-70569 Stuttgart, Germany\\
         and Institut f\"ur Theoretische und Angewandte Physik, 
              Universit\"at Stuttgart,\\D-70550 Stuttgart, Germany} 
\date{\today}
\maketitle
\sloppy
\begin{abstract}

Fluids consisting of hard platelike particles near a hard wall
are investigated using density functional theory. The density and orientational 
profiles as well as the surface tension and the excess coverage are determined and 
compared with those of a fluid of rodlike particles. Even for 
low densities slight orientational packing effects are found for the platelet 
fluid due to larger intermolecular interactions between platelets as 
compared with those between rods. A net depletion of platelets near the wall is 
exhibited by the excess coverage, whereas a change of sign of the excess coverage 
of hard-rod fluids is found upon increasing the bulk density.
\end{abstract}
\pacs{PACS numbers: 61.20.-p, 61.30.Gd, 82.70.Dd}
}
\narrowtext

\section{Introduction}  
While many theoretical studies have been devoted to the understanding 
of the behavior of elongated hard colloidal particles near a hard wall 
(see, e.g., Refs.~\cite{poni:88}-\cite{dijk:01}),
suspensions of disc shaped hard particles near a hard wall have not yet been 
investigated. One reason is that experimentally a corresponding hard platelet 
model system, i.e., consisting of particles with a short range repulsive interaction 
has been lacking until recently. From a theoretical point of view the platelet 
fluid problem appears to be difficult because the Onsager approach 
of truncating the corresponding equation of state at second order, which is 
valid for thin rodlike particles, cannot be justified for platelets \cite{onsa:49}. 
Recently, preparation methods have been developed for new types of platelet suspensions, 
which may indeed serve as model systems of hard colloidal platelets 
\cite{brow:98,kroj:98,brow:99}. In the present article we use density functional theory 
(Sec. II) to study the positional and orientational order as well as the surface 
tension and excess coverage of thin hard platelets near a hard wall (Sec. III). 
Particularly, the density functional used here includes a third order density term 
which is not present in the Onsager second virial approximation.

\section{Model and density functional theory}  
We consider an inhomogeneous fluid consisting of thin platelets of radius $R$
in a container of volume $V$. The platelets are taken to be hard discs without 
attractive interactions. The number density of the centers of mass of the
platelets at a point ${\bf r}$ with an orientation $\omega=(\theta,\phi)$ of 
the normal of the platelets (see Fig.~\ref{fig1}) is denoted by $\rho({\bf r},\omega)$. 
The equilibrium density profile of the inhomogeneous 
liquid under the influence of an external potential $V({\bf r},\omega)$ minimizes 
the grand potential functional \cite{evan:92}:

\begin{figure}[h] 
\vspace*{-1.0cm}
\hspace*{2.0cm}
\epsfxsize=3.5cm
\epsffile{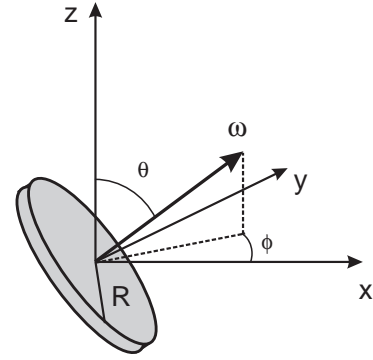}
\vspace*{1.15cm}
\caption{The polar angle $\theta$ and the azimuthal angle $\phi$ of the normal 
\mbox{\boldmath$\omega$} of a thin platelet of radius $R$ with its center of mass located 
at ${\bf r}=(0,0,0)$.}
\label{fig1}
\end{figure}

\begin{eqnarray} \label{eq1}
\Omega[\rho({\bf r},\omega)]&=&\int d{\bf r}\,d\omega\,\rho({\bf r},\omega)
\left[k_BT\left(\ln(4\pi\Lambda^3\rho({\bf r},\omega))-1\right)\right.\nonumber
\\&&-\left.\mu+ V({\bf r},\omega)\right]+ F_{ex}[\rho({\bf r},\omega)]\,,
\end{eqnarray}
where $\Lambda$ is the thermal de Broglie wavelength and $\mu$ is the 
chemical potential. We express the excess free energy functional 
$F_{ex}[\rho({\bf r},\omega)]$  as an integral over all spatial and 
orientational degrees of freedom for the corresponding local free energy 
density $f_{ex}(\rho=\rho({\bf r},\omega))$ of the fluid:
\begin{eqnarray} \label{eq2}
F_{ex}[\rho({\bf r},\omega)]&=&\int d{\bf r}_1\,d\omega_1\,d{\bf r}_2\,d\omega_2\,
w({\bf r}_1,{\bf r}_2,\omega_1,\omega_2)\nonumber
\\&&\times\rho({\bf r}_1,\omega_1)f_{ex}(\rho({\bf r}_2,\omega_2))\,,
\end{eqnarray}
with
\begin{equation} \label{eq2a}
f_{ex}(\rho)=4\pi^2\left[\sqrt{2} R^3\rho
+\frac{4\pi^2}{3}R^6\rho^2\right]k_BT\,.
\end{equation}
This functional incorporates fluid correlations via a weight function
$w({\bf r}_1,{\bf r}_2,\omega_1,\omega_2)$ and
bulk thermodynamics via a recently developed 
equation of state of thin platelets  \cite{harn:01}:

\begin{equation} \label{eq3}
p_b=-\left(\frac{\partial \Omega}{\partial V}\right)_{T,\mu}=
\rho_b\left(1+\sqrt{2}\pi R^3\rho_b+\frac{2\pi^2}{3}R^6\rho_b^2
\right)k_BT\,,
\end{equation}
where $\rho_b=V^{-1}\int dr\,\int d\omega\, \rho({\bf r},\omega)$
is the number density of the homogeneous and isotropic bulk fluid.
We note that the resulting  excess free energy functional 
$F_{ex}[\rho({\bf r},\omega)]$ includes a cubic term of the density, 
which is not present in the Onsager second virial 
\mbox{approximation \cite{onsa:49} }
\begin{eqnarray} \label{eq3a}
f_{ex}^{({\rm rod})}(\rho)=\pi^2DL^2\rho k_BT
\end{eqnarray}
used in the description of thin rods of length $L$ and diameter $D$ near surfaces 
(see, e.g., Refs. \cite{poni:93,mao:97} and \cite{groh:99}). 
The necessity for including this higher order density term already follows 
from comparing the ensuing bulk pressure (Eq. (\ref{eq3})) with corresponding 
simulation data \cite{eppe:84,dijk:97} (see Fig. \ref{fig2}). 
Fluids consisting of thin platelets exhibit an isotropic
and a nematic phase with no other liquid crystalline phases, such as a 
columnar phase, observed \cite{eppe:84}. The isotropic-nematic (IN) transition 
is first order with coexistence densities at 
$\rho_{bI}R^3=0.46$ and $\rho_{bN}R^3=0.5$ \cite{bate:99}.

Minimization of $\Omega$ with respect to $\rho({\bf r},\omega)$
leads to the following integral equation for the equilibrium density 
distribution:
\begin{eqnarray} \label{eq6}
&&k_BT\ln[4\pi\Lambda^3\rho({\bf r},\omega)]=\mu- V({\bf r},\omega)
\nonumber
\\&&-8\pi^2 R^3\int d{\bf r}_1\,d\omega_1\rho({\bf r}_1,\omega_1)
w({\bf r},{\bf r}_1,\omega,\omega_1)\nonumber
\\&&\times\left[\sqrt{2}+\frac{2}{3}\pi^2 R^3[2\rho({\bf r},\omega)
+\rho({\bf r}_1,\omega_1)]\right]k_BT\,.
\end{eqnarray}
We have solved this equation numerically for a given chemical potential 
$\mu$ and a given external field $V({\bf r},\omega)$.
The weight function is taken to be a function of the relative positions
${\bf r}_{12}={\bf r}_1-{\bf r}_2$, and is normalized so that 
$\int d{\bf r}_{12}\,d\omega_2\,w({\bf r}_{12},\omega_1,\omega_2)=1$.
In this work the weight function is given by the Mayer function, except for the 
normalization. The Mayer function equals -1 if the platelets overlap 
and is zero otherwise. 
Two platelets, separated by a distance ${\bf r}_{12}$, 
intersect if the inequality 
\begin{eqnarray} \label{eq9}
|{\bf r}_{12}\cdot (\mbox{\boldmath$\omega$}_1\times
\mbox{\boldmath$\omega$}_2)|&<&
\sqrt{R^2\sin\gamma_{12}^2-({\bf r}_{12}\cdot 
\mbox{\boldmath$\omega$}_2)^2}\nonumber
\\&&+\sqrt{R^2\sin\gamma_{12}^2-({\bf r}_{12}\cdot 
\mbox{\boldmath$\omega$}_1)^2}\,,
\end{eqnarray}
with 
$\mbox{\boldmath$\omega$}_j=(\sin\theta_j\cos\phi_j,\sin\theta_j\sin\phi_j,\cos\theta_j),
\hspace{0.5cm} j=1,2$,
is fulfilled. $\gamma_{12}$ is the angle between the normals 
$\mbox{\boldmath$\omega$}_1$ and $\mbox{\boldmath$\omega$}_2$ of two platelets. 
Equation (\ref{eq9}) has been derived and is amply documented in Ref.~\cite{eppe:84}. 
In practice, first the  weight function is calculated by testing Eq.~(\ref{eq9}) 
and is stored for all required values of (${\bf r}_{12},\omega_1,\omega_2$).

\begin{figure}[h]   
\vspace*{-1.0cm}  
\epsfysize=12cm  
\epsffile{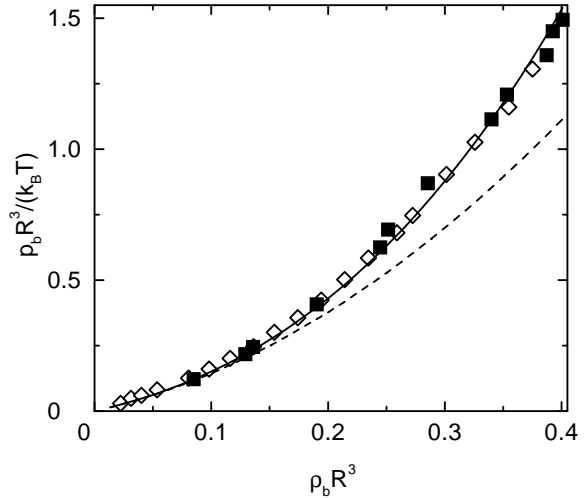}
\vspace*{-3.5cm}  
\caption{Equation of state of thin platelets as obtained from   
Eq.~(\protect{\ref{eq3}}) (solid line), and from  computer   
simulations (squares \protect\cite{eppe:84} and diamonds \protect\cite{dijk:97}).
The dashed line follows from Eq.~(\protect{\ref{eq3}}) by omitting the cubic 
term $\sim \rho_b^3$.}  
\label{fig2}   
\end{figure}

Thereafter the integral equation (\ref{eq6}) is solved using a Picard scheme 
with retardation.

\section{Platelets and rods near a hard wall} 
\begin{figure}[h] 
\vspace*{-0.6cm}  
\epsfysize=7.5cm  
\epsffile{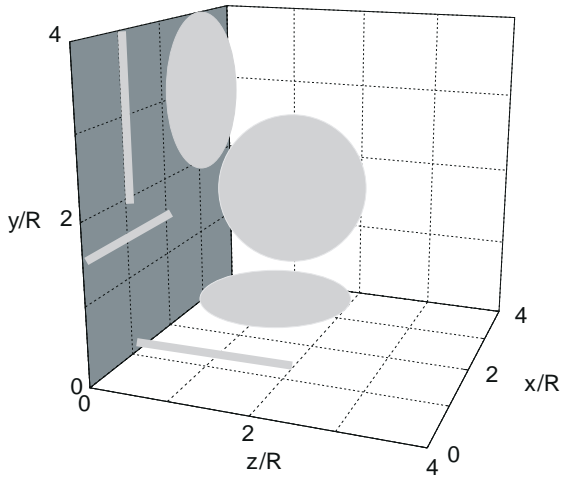}
\vspace*{1cm}   
\caption{The two systems under consideration consist of fluids of monodisperse thin 
platelets of radius $R$ and thin rods of length $L=2R$ in contact with a planar hard 
wall at $z=0$. Particles very close to the wall must adopt nearly a fully parallel 
alignment. Here the three principal directions are shown.}  
\label{fig3} 
\end{figure}

For model systems of hard particles near a hard wall at \mbox{$z=0$} (see Fig.~\ref{fig3}),
apart from a possible surface freezing at high densities, nonuniformities of the 
density occur only in the $z$ direction, so that
$\rho({\bf r},\omega)=\rho(z,\theta,\phi)$. Figure \ref{fig4} displays the 
calculated platelet density profile for the bulk density $\rho_bR^3=0.2$.
The calculations render $\rho(z,\theta,\phi)$ to be independent of the azimuthal 
angle $\phi$ for this density, i.e., there is no biaxial order emerging at the 
wall like for hard rods at high densities \cite{roij:00a,roij:00b,dijk:01}. 
For $z<R$ orientations with large $\theta$
values are forbidden so that the density profile has a discontinuity along the 
line $z_{min}=R \sin\theta$. There is a pronounced increase of the density near 
the surface. Moreover, slight packing effects are visible through oscillations at 
$z\approx 2R$. For comparison Fig.~\ref{fig5} displays the density profile of thin 
rods near a hard wall calculated from Eqs.~(\ref{eq1}), (\ref{eq2}) and (\ref{eq3a}). 
The bulk density has been fixed such that the 
second virial coefficient of the equation of state of thin platelets 
(see Eq.~(\ref{eq3})) and of the equation of state of thin rods, 
\begin{equation} \label{eq10}
p_b^{(\rm rod)}=\rho_b\left(1+\frac{\pi}{4}DL^2\rho_b\right)k_BT\,,
\end{equation}
are equal in units of $R^3$ and $DL^2$, respectively. 
Due to the presence of the wall $\rho(z,\theta,\phi)$ vanishes if 
\mbox{$z<z_{min}=(L/2)\cos\theta$.}

Hence, in terms of the normals of the platelets 
and the normals of the rods along their main axis of symmetry the preferred 
orientation of rods near a hard wall 
is perpendicular to the preferential
orientation  of the platelets. This means that in both cases the main body 
of the particles tends to lie parallel to the wall (see Fig.~\ref{fig3}). 
In agreement with earlier calculations \cite{groh:99}, for rods no packing 
effects are visible because of the relatively smaller intermolecular 
interactions between rods as compared with those between platelets.
(For thin rods the intersection volume is pointlike whereas for discs
it is like a line segment.)

We have tested successfully the accuracy of the numerical calculations by 
comparing our results for the profiles with the  pressure sum rule 
\begin{equation} \label{eq11}
p_b=k_BT\int\limits_0^{2\pi} d\phi\,\int\limits_0^{\pi} d\theta\,
\sin\theta\,\rho(z_{min},\theta,\phi)\,,
\end{equation}
and we found good agreement. Equation (\ref{eq11}) is an extension of the 
pressure sum rule of a hard-sphere fluid near a wall 
(see, e.g., Ref. \cite{swol:89}). The weighted density functional theory 
guarantees that the wall sum rule is satisfied, provided that the pressure 
enters through the corresponding bulk equation of state. 

A set of position-dependent order parameters quantifies the deviation
from isotropy of the number density. For a uniaxial density profile the
most general form of $\rho(z,\theta,\phi)$ is independent of $\phi$ so 
that one can write:

\begin{figure}[h] 
\vspace*{0.25cm}  
\epsfysize=7cm  
\epsffile{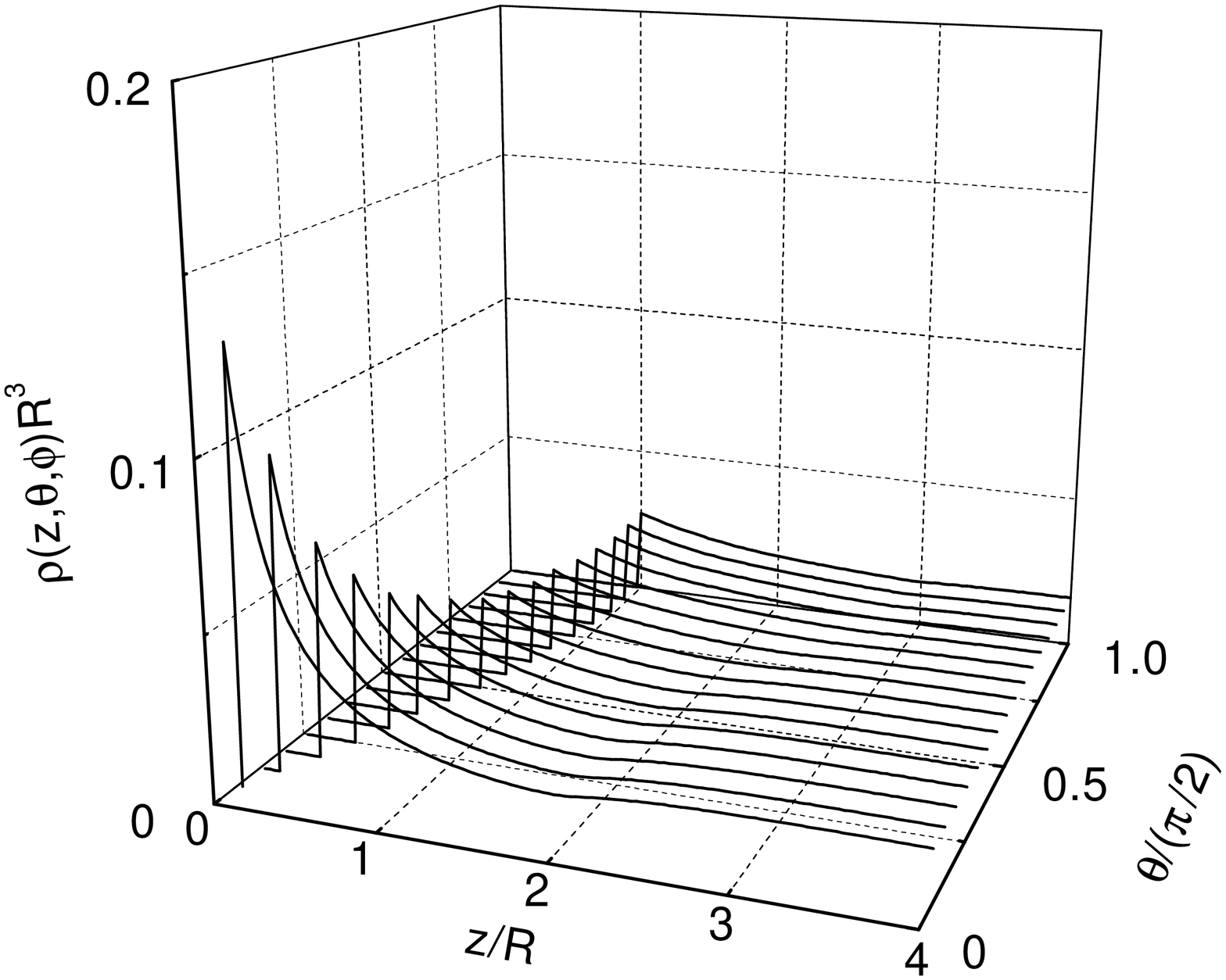}
\vspace*{0.5cm}  
\caption{Center of mass density profile $\rho(z,\theta,\phi)R^3$ 
of thin platelets near a hard wall for various angles $\theta$ of the platelet 
normals relative to the surface normal. The density profile is independent of the 
azimuthal angle $\phi$. At small distances $z$ from the wall, large values of
$\theta$ are forbidden due to overlap. Therefore the profile is exactly zero 
if $z<z_{min}=R\sin\theta$. The bulk density is fixed to $\rho_bR^3=0.2$.}  
\label{fig4}   
\end{figure}

\begin{figure}[h]   
\vspace*{-0.3cm}  
\epsfysize=7cm  
\epsffile{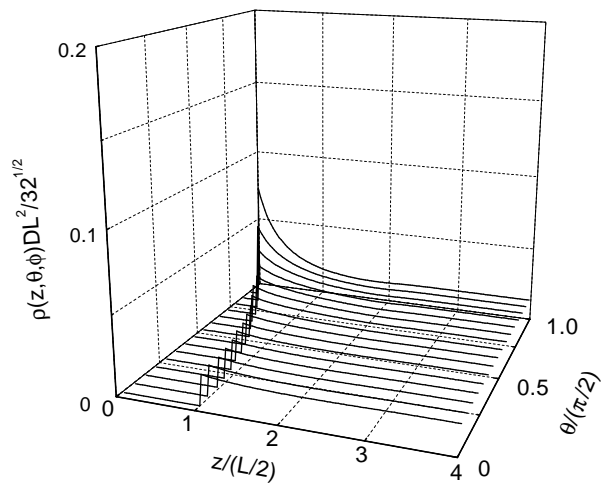}
\vspace*{0.5cm}  
\caption{Center of mass density profile $\rho(z,\theta,\phi)DL^2/\sqrt{32}$ 
of thin rods ($L/D\to \infty$) near a hard wall for the bulk density 
$\rho_bDL^2/\sqrt{32}=0.2$. The profile is exactly zero if $z<z_{min}=(L/2)\cos\theta$ 
due to overlap.}  
\label{fig5}   
\end{figure}

\begin{equation} \label{eq12}
\rho(z,\theta)=\sum\limits_{l=0}^\infty \frac{2l+1}{2}Q_l(z)P_l(\cos(\theta))\,,
\end{equation}
where $P_l(\cos(\theta))$ are the Legendre polynomials. 
The normalized, orientationally averaged number density profiles

\begin{figure}[h]   
\vspace*{-0.3cm}  
\epsfysize=12cm  
\epsffile{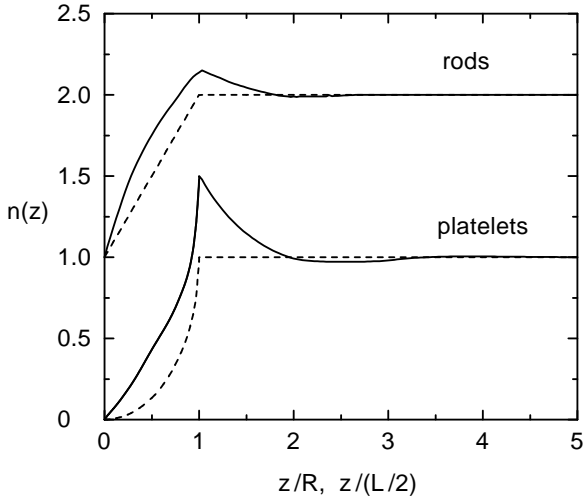}
\vspace*{-3.8cm}  
\caption{Normalized, orientationally averaged density profile $n(z)$
for platelets (lower solid curve) and rods (upper solid curve) in contact 
with a hard wall for the same bulk density as in Figs.~\protect\ref{fig4} and \ref{fig5}. 
The length $L$ of the rods and the diameter of the platelets $2R$ are taken 
to be the same. The dashed lines represent the results for an ideal gas 
of platelets and rods, respectively. The upper curves are shifted by 1.}  
\label{fig6}   
\end{figure}

$n(z)=2\pi Q_0(z)/\rho_b$ 
are displayed in Fig.~\ref{fig6} together with the results for non-interacting 
platelets and rods. Upon increasing $z$ from the wall the averaged number densities 
increase and exhibit cusps at $z=R$ for platelets, and at $z=L/2$ for rods, respectively. 
For rods the 
maximum at the cusp is only about 10\% above the bulk value which is essentially
reached already for $z=L$. For platelets the maximum is more pronounced and slight 
packing effects are visible at larger values of $z$ due to the relatively larger steric 
interactions between platelets.

\begin{figure}[h]   
\vspace*{-0.3cm}  
\epsfysize=12cm  
\epsffile{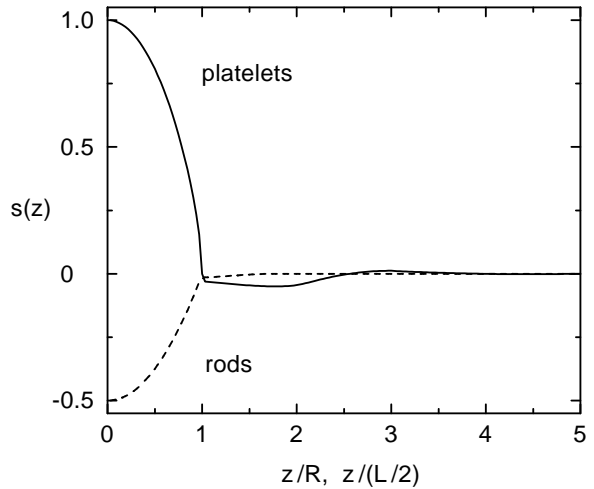}
\vspace*{-3.8cm}  
\caption{Relative uniaxial nematic order parameter $s(z)$ for platelets  
(solid curve) and rods (dashed curve) in contact with a hard wall for the same bulk 
density as in Figs.~\ref{fig4} and \ref{fig5}. Positive (negative) values of the 
nematic order parameter indicate that the platelets (rods) are preferentially 
aligned parallel to the wall.}  
\label{fig7}   
\end{figure}

The position-dependent uniaxial, relative nematic order parameter 
$s(z)=Q_2(z)\rho_b/(2\pi Q_0(z))$
is displayed in Fig.~\ref{fig7}. At small values of $z$ the value of the nematic order 
parameter reflects the geometric constraints. 
A platelet [rod] lying very closely to the wall must adopt 
nearly a fully parallel alignment, 
so that the order parameter reaches the limiting value  $1=P_2(\cos(0))$,
[$-0.5=P_2(\cos(\pi/2))$] there, whereas 
isotropic orientation at large distances from the wall is characterized by 
$s(z)=0$. Interestingly, the nematic order parameter for platelets
has a minimum at $z\approx 2R$  due to a depletion of platelets parallel to the 
wall. In other words, a platelet, next to the platelet at $z=R$ with the rim touching 
the wall (see Fig. \ref{fig3}), is oriented rather parallel than 
perpendicular to the former. 

It is useful to consider not only local but also global properties of a liquid 
near a surface. There are two global quantities, which are of experimental interest
and of interest for simulations:
the excess coverage, which is

\begin{figure}[h]   
\vspace*{-0.3cm}  
\epsfysize=12cm  
\epsffile{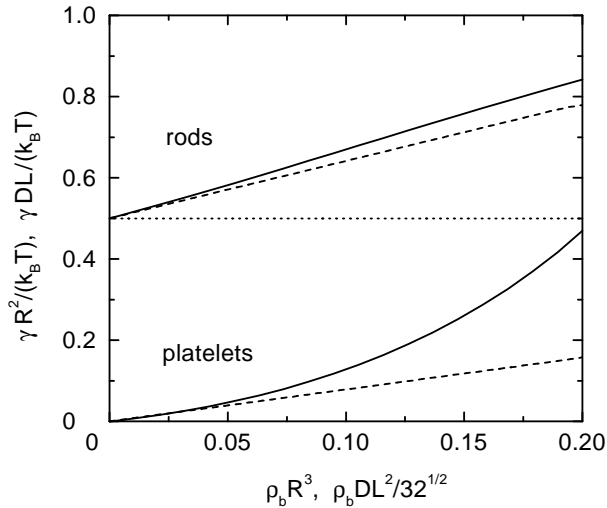}
\vspace*{-3.8cm}  
\caption{Surface tension of a fluid consisting of thin hard platelets of radius $R$
(lower solid curve) and hard rods of length $L$ and diameter $D$ (upper solid curve, 
$L/D \to \infty$) near a hard wall.
The dashed lines represent the results for an ideal gas of platelets and rods, 
respectively. The upper curves are shifted by 0.5 (dotted line).}  \label{fig8}   
\end{figure}

accessible by, e.g., gravimetric measurements;
and the surface tension which, for example, is important for contact 
angles. The surface tension  $\gamma$ is defined via
\begin{equation} \label{eq13}
\Omega[\rho(z,\theta,\phi)]=V\omega_b+\gamma S\,
\end{equation}
where $S$ is the surface area and $\omega_b=-p_b$ (see Eq. (\ref{eq3})) is the bulk 
grand-canonical potential density. The surface tension depends on the definition 
of what is denoted as the volume $V$ \cite{lajt:87}. We have defined $V$ as the 
volume of the container with its surface given by the 
position of the rim of the particles at closest approach so that the region 
$0<z<z_{min}$ is part of $V$. Figure \ref{fig8} displays the 
calculated surface tension together with the results for non-interacting platelets and rods. 
The steric interaction between the particles, which is more pronounced for the platelets, 
increases the surface tension with increasing density. 
The results for the rods are in agreement with those obtained in Refs. \cite{mao:97}
and \cite{groh:99}.

\begin{figure}[h]   
\vspace*{-0.3cm}  
\epsfysize=12cm  
\epsffile{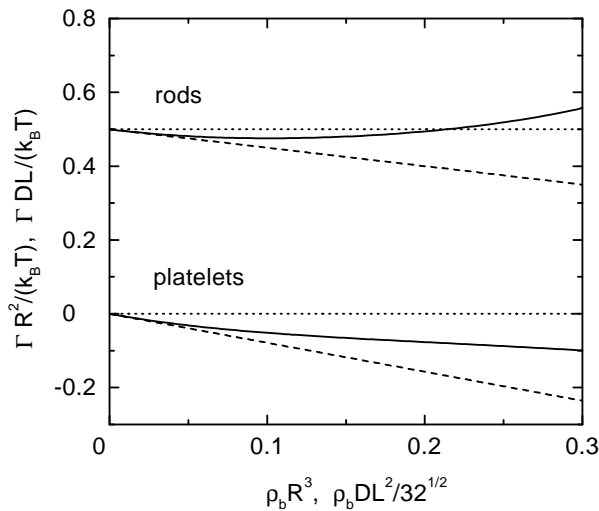}
\vspace*{-3.8cm}  
\caption{Excess coverage (see Eq.~(\ref{eq14})) of a fluid consisting of thin platelets 
of radius $R$ (lower solid curve) and rods of length $L$ and diameter $D$
(upper solid curve, $L/D \to \infty$) near a hard wall.
The dashed lines denote the corresponding results for an ideal gas of platelets and rods, 
respectively. The upper curves are shifted by 0.5.}  \label{fig9}   
\end{figure}

The excess coverage 
\begin{equation} \label{eq14}
\Gamma=\rho_b\int\limits_0^\infty dz\,\left(n(z)-1\right)
\end{equation}
provides an important overall characterization of the density profiles. Figure \ref{fig9} 
summarizes the results for platelets and rods. Repulsive interactions between the platelets
and the hard wall lead to a net depletion near the surface ($\Gamma<0$). 
The  excess coverage 
of rods exhibits a change of sign with increasing density in agreement with

recent density functional calculations \cite{roij:00a,roij:00b} and computer 
simulations \cite{dijk:01}.

\section{Summary}
We have applied a density functional theory to fluids consisting of thin hard platelets
and rods in contact with a hard wall. Particles lying very close to the wall must adopt 
nearly a fully parallel alignment due to interactions with the wall (Fig.~\ref{fig3}).
The probability of finding a particle touching the wall is increased 
compared with the bulk (Figs.~\ref{fig4} and \ref{fig5}). A comparison between the rod fluid 
and the platelet fluid exhibits slight orientational packing effects for the platelet 
fluid (Fig.~\ref{fig7}) and that the increase of the surface tension 
with increasing density is more pronounced for platelets than for rods (Fig.~\ref{fig8}) 
due to larger intermolecular interactions between platelets as 
compared with those between rods. The calculated excess coverage (Fig.~\ref{fig9})
reveals a depletion of platelets close to the wall (Fig.~\ref{fig6}), whereas a change
of sign of the excess coverage of hard-rod fluids indicate the onset of wetting.

\acknowledgments
L. H. gratefully acknowledges support by the Deutsche Forschungsgemeinschaft.

\end{document}